\documentclass[aps,prl,reprint,groupedaddress]{revtex4-1}

\usepackage{amsmath}
\usepackage{graphicx}
\usepackage{verbatim}
\usepackage{xspace}
\usepackage{xcolor}

\begin{document}

%\preprint{}

\title{Watch and learn -- a generalized approach for transferrable learning in deep neural networks via physical principles}

%\title{Unraveling interactions -- a generalized approach for transferrable learning in deep neural networks via physical principles}

%\title{Found in translation -- a generalized approach for transferrable learning in deep neural networks via physical principles}



\author{Kyle Sprague}
\affiliation{Department of Physics, University of Ottawa, Ontario, Canada}

\author{Juan Carrasquilla}
\affiliation{Vector Institute, Toronto, Ontario, Canada}
\affiliation{Department of Physics and Astronomy, University of Waterloo, Ontario, Canada}

\author{Steve Whitelam}
\affiliation{Molecular Foundry, Lawrence Berkeley National Laboratory, Berkeley, California, USA}

\author{Isaac Tamblyn}
\email{isaac.tamblyn@nrc-cnrc.gc.ca}
\affiliation{National Research Council of Canada, Ottawa, Canada}
\affiliation{Vector Institute, Toronto, Ontario, Canada}
\affiliation{Department of Physics, University of Ottawa, Canada}

\date{\today}

\begin{abstract}

Transfer learning refers to the use of knowledge gained while solving a machine learning task and applying it to the solution of a closely related problem. Such an approach has enabled scientific breakthroughs in computer vision and natural language processing where the weights learned in state-of-the-art models can be used to initialize models for other tasks which dramatically improve their performance and save computational time. Here we demonstrate an unsupervised learning approach augmented with basic physical principles that achieves fully transferrable learning for problems in statistical physics across different physical regimes. By coupling a sequence model based on a recurrent neural network to an extensive deep neural network, we are able to learn the equilibrium probability distributions and inter-particle interaction models of classical statistical mechanical systems. Our approach, distribution-consistent learning, DCL, is a general strategy that works for a variety of canonical statistical mechanical models (Ising and Potts) as well as disordered (spin-glass) interaction potentials. Using data collected from a single set of observation conditions, DCL successfully extrapolates across all temperatures, thermodynamic phases, and can be applied to different length-scales. This constitutes a fully transferrable physics-based learning in a generalizable approach.

\end{abstract}


\maketitle

\section{Introduction}

Machine learning has emerged as a powerful tool in the physical sciences, seeing both rapid adoption and experimentation in recent years. Already, there have been demonstrations of learning operators~\cite{mills_pre_2018}, detecting unseen patterns within data~\cite{Chng2016a}, ``discovering'' physical equations~\cite{WANG20191228}, and predicting trends within the scientific literature~\cite{tshitoyan_nature_2019}. Within the field of statistical mechanics, machine learning was recently used to estimate the value of the partition function~\cite{desgranges2018new}, solve canonical models~\cite{wu_prl_2019}, and generative models conditioned on the Boltzmann distribution have been shown to be efficient at sampling statistical mechanical ensembles\cite{noe_boltzmann_2019}. The connection between physics and machine learning continues to strengthen, with new results appearing daily.

Despite these and other successes, machine learning has severe limitations. A major obstacle to the more widespread use of machine learning in the physical sciences is that typically, learned models tend to exhibit poor transferability. Using a model outside of the training or parameterization set can be unreliable~\cite{goodfellow_book}. This, along with the lack of well defined and well behaved error estimates can result in erroneous results, the magnitude of which are often uncontrolled. While transferability can be somewhat improved through techniques such as regularization~\cite{dropout}, there is currently no general approach which can guarantee transferability to new conditions or ensure reliability of a model when it is presented with previously unseen data.

Here we demonstrate a new approach which overcomes the transferability problem by coupling machine learning concepts with physical principles in a new way. The approach, which we call distribution-consistent learning (DCL) enables fully transferrable learning with a minimal number of observations: using it, we can collect observations at a single set of conditions, yet make accurate predictions for all others (including in different physical phases and across phase boundaries). Using DCL, it is possible to extrapolate observations over a wide range of conditions, including those far from the training set. This is achieved through the straightforward yet rigorous application of the physical concept of equilibrium and the postulate of the uniformity of physical law.

To explain DCL, we will first focus on simple classical spin models (ferromagnetic Ising and Potts models with coupling constant $J=1$) for the purposes of illustration and validation.  For such simple cases, application of statistical methods to ``invert'' observations are not new~\cite{nguyen_aip_2017, valleti_arxiv_2019}. 

Conceptually, DCL is valid when ensemble probabilities are governed by a known statistics (e.g. Boltzmann or Fermi distributions, which describe equilibrium and some near-equilibrium processes) and the interaction energies do not depend on the control parameters.

Finally, we demonstrate the generalizability of DCL explicitly by applying it, without modification, to an unsolved inversion problem: a semi-local ``spin-glass'' where all couplings between neighbours are selected randomly from a Gaussian distribution ($\mu=0$, $\sigma=1$).

\section{Probabilities at equilibrium}

When a system is at equilibrium, its macroscopic properties can be interpreted as weighted averages over the character of a large number of micro-states, ${\sigma_i}$. For an example such as a polymer, such micro-states may be different molecular conformations, whereas for a spin-system they are the different combinations of possible up (down) spin arrangements. Such micro-states are visited with a probability given by the Boltzmann distribution. For the case of the canonical ensemble in particular (constant particle number, $N$, volume, $V$, and temperature, $T$), the ensemble probability of a particular micro-state is given as

\begin{equation}
P(\sigma_i) = \exp(-\beta \mathcal H(\sigma_i))/Z,
\label{eqn:probability}
\end{equation}

where $\mathcal H$ is the classical Hamiltonian (and the energy, $E_i$, of $\sigma_i$ is given by $\mathcal H(\sigma_i)$), $\beta$ is the inverse temperature, $\frac{1}{k_B T}$, and $Z$ is the partition function.

We note that in principle if one were able to observe an equilibrium system long enough, it would be possible to estimate the ensemble probabilities of each micro-state simply by counting -- count how often the system visits a particular micro-state and divide by the number of observed events. This information, combined with the observation temperature ($\beta_O$), and Eqn.~\ref{eqn:probability} would be sufficient to determine the energy differences between any two micro-states $\sigma_i, \sigma_j$

\begin{equation}
\Delta E_{ij} = \beta_O^{-1} \ln\frac{P(\sigma_j)}{P(\sigma_i)}.
\label{eqn:delta_e}
\end{equation}

Of course, for all but trivial state spaces, such an approach is completely impractical. The probability of visiting micro-states above the ground state at finite $T$ is exponentially small in $E$, and the probability of visiting them multiple times within an observation period (which would be necessary in order to collect reliable estimates) is even smaller. Micro-states visited during a macroscopic experiment represent only a vanishingly small fraction of the possible configurations which can be realized. Watching and counting is not an option.

\section{Essence of the approach}

\begin{figure}[t]
 \includegraphics[width=\columnwidth]{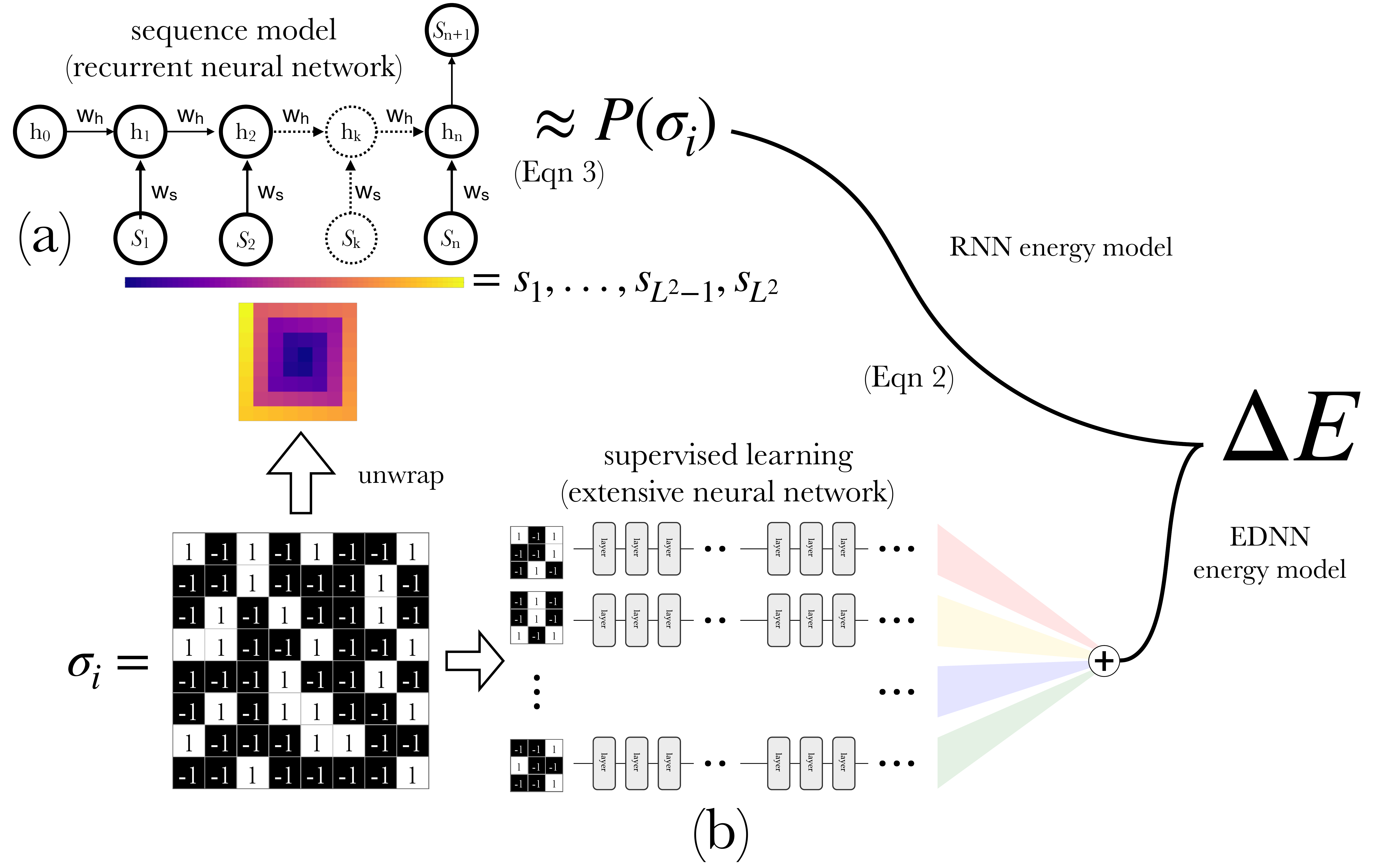}
 \caption{Distribution-consistent learning: we conduct numerical experiments at a single set of ``experimental'' conditions, observe the visited micro-states, $\sigma_i$, and train a sequence model, a), (here a recurrent neural network, RNN) to learn their structure. RNN probability predictions can train a second neural network, b), via supervised learning. The estimate of $P(\sigma_i)$ is obtained from the RNN through a chain of conditional probabilities (Eqn~\ref{eqn:pi_product}). Applying the same process to a second configuration $\sigma_j$ results in an estimate of the energy difference of the two, $\Delta E$. \label{fig:fig1_workflow}}.
\end{figure}

Rather than attempt to infer probabilities from counting, we instead consider the structure of $\sigma$ itself. It is generally known that interactions between spins can give rise to correlations on a range of scales, and that emergence of such correlations are temperature dependent. In some sense, these correlations are similar to patterns that emerge in language due to the rules of grammar. We can ``unwrap'' $\sigma$ into a sequence (Fig.~\ref{fig:fig1_workflow}) and analyze them using language-based sequential machine learning methods. While many such sequence models exist, recurrent neural networks (RNN) are a particularly powerful deep learning technique which have seen significant use in recent years. Primarily, RNN have been used in natural language processing (e.g. predictive text and translation), as well as predicting time-series data such as stock markets and weather. More recently, autoregressive RNN have been adapted to spatially structured inputs such as images~\cite{pixelRNN}.

We first train an RNN on the micro-states observed at a fixed set of experimental temperature conditions, $\beta_O^{-1}$ (for this study we used $\beta_O^{-1}=4$) as DCL requires only data collected at a single temperature. Formally, the precise temperature of observation does not matter, so long as the system is approximately ergodic. Here we tested this explicitly by increasing the temperature by a factor of $\times2$. This did not change our results qualitatively. 
 
Our training set consists of observations ${\sigma_i}$ as well as the thermodynamic conditions upon which they were collected (i.e. the observation temperature $\beta_O^{-1}$). Importantly, the Hamiltonian operator and values of the energy are \emph{not included} in our training data. This is analogous to what occurs experimentally - one does not generally have knowledge of the interaction potential between particles, but can always conduct an experiment at fixed conditions and make observations of which micro-states are explored during such an experiment.

To train the RNN, we provided it with examples of spin micro-states as unwrapped configurations (Fig. 1~\ref{fig:fig1_workflow}). We experimented with different ``unwrapping'' techniques (which we called ``snake'' or ``spiral'' respectively); tests confirm our results are not sensitive to this choice, so long as the procedure is used consistently. Our training procedure and network architecture follow standard techniques and are reported as Supplementary Information.

Once the RNN has been trained, we can use it to determine a particular $P(\sigma_i)$ by computing a series of conditional probabilities of the unwrapped $\sigma_i$ sequence. To do this, we initialize the RNN with the first spin value of $\sigma_i[1] = s_1$ and record its prediction for $P(s_1)$. Applying this now initialized RNN model to $s_2$ produces a conditional probability prediction $P(s_2 | s_1)$. The total probability of $P(\sigma_i)$ is found from the multiplicative product of many conditional probabilities

\begin{equation}
P(\sigma_i) = P(s_1)\prod^{L^d}_{i=2}{P(s_{i} | s_{i - 1}...s_1)},
\label{eqn:pi_product}
\end{equation}

where $L$ is the linear dimension of the system and $d$ is the dimension.

For a two state system (Ising), the RNN has only a single output since $P(\text{up}) = 1-P(\text{down})$. Ising has an input vector of size one (for the spin) and an equivalent output vector. We generalize this for the Potts model, which has an input vector of size $Q$ (either a 1 or a 0 in each element). The output is also a $Q$ wide vector.

% This is only for a particular case, not in general. $P(s_1)=1/Q$, and $Q$ is the number of possible states (e.g. for Ising, $Q=2$).

Using Eqn.~\ref{eqn:pi_product}, we can compare the probabilities of any two micro-states, $P(\sigma_i)$, $P(\sigma_j)$. Inserting these into Eqn.~\ref{eqn:delta_e} (combined with the experimental conditions under which the samples were obtained, $\beta_O$)) provides us with an estimate of the difference of internal energy between the micro-states $\sigma_i, \sigma_j$. We are able to compute such an estimate for any pair, including configurations which were never visited during the training process. Henceforth, we refer to this process of estimating $E_i$ as the RNN energy model.

\section{Using the RNN Energy Model}

\begin{figure}[!h]
 \includegraphics[width=\columnwidth]{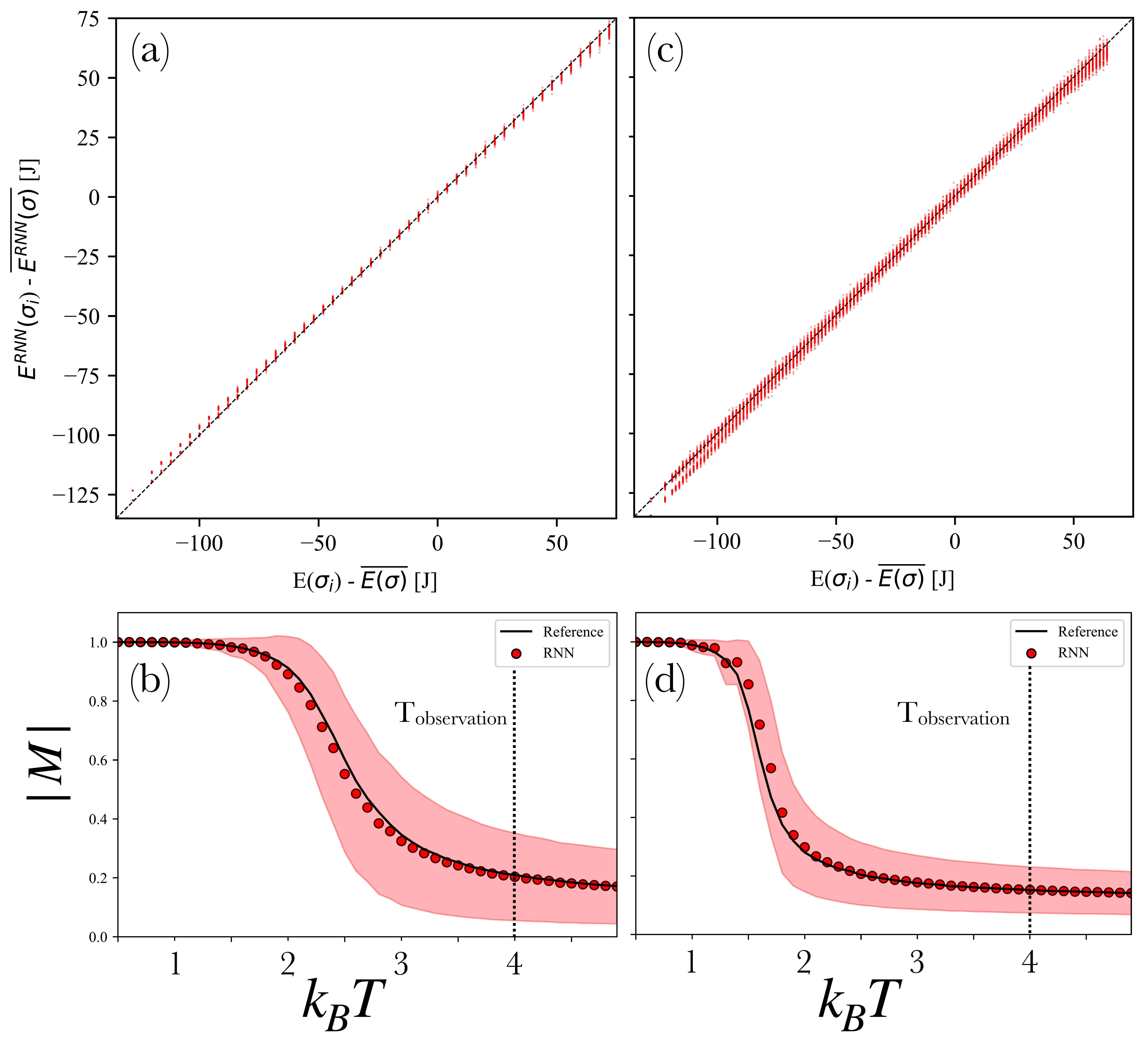}
 \caption{Performance of RNN energy models for the cases of Ising (a) and Potts (b) compared to their true values evaluated over a wide range of energies (we use the average energy of the dataset as a reference energy $\overline{E^{\text{RNN}}(\sigma)}$ for each model). The black diagonal line is a guide to the eye. Using these models of the energy with Metropolis Monte Carlo, it is possible to predict properties over a wide range of temperatures (bottom row, b and d). We note that RNN energy models are trained only on samples which are recorded at $T_o$ (denoted as a vertical dashed line), yet can be used to make predictions across a wide range of physical conditions, including through a phase transition. The red shaded area indicates the variance observed in our numerical simulations.\label{fig:rnn_transition}}
\end{figure}

The top row of Fig.~\ref{fig:rnn_transition} show the performance of the RNN energy models across a range of energies for both the Ising and Potts models. In both cases, the RNN energy model does an excellent job estimating the energy difference between any two micro-states. We note that it is exactly this quantity which is needed to perform finite temperature Metropolis Monte Carlo simulations to predict the thermodynamic properties of a spin system.

Using these models, we carry out such numerical simulations under conditions which are far away from the training set. When we compare results for the phase transitions generated using the true Hamiltonian with those generated with the RNN energy models, we see excellent agreement for both Ising and Potts models (Fig.~\ref{fig:rnn_transition} bottom row). This confirms that the RNN model errors in the prediction of $\Delta E_{ij}$ are small enough that they do not alter the essential physics of systems under study.

Since the RNN already allows us to estimate ensemble probabilities (and thus energy labels from observation), we might think that nothing more can be extracted from our initial observations. It turns out, however, we can incorporate more physics knowledge through the use of a second neural network. We will demonstrate that incorporating this second network will both overcome some limitations of the RNN energy model and improve the accuracy of our predictions \emph{without any new observations or labels}.

One of the most obvious disadvantages of the RNN energy model is that it has no concept of locality with respect to energy updates. Whenever we wish to know the difference in energy between $\sigma_i$ and $\sigma_j$, we must reevaluate all of the conditional probabilities which make up $P(\sigma_i)$ and $P(\sigma_j)$. For interactions which are short ranged (as is the case for both the Ising and Potts models), this is very inefficient. In conventional simulations of such systems, it is customary to reevaluate only interactions which have changed as a result of the MC trial move. With the RNN energy model, there is no general way to achieve such ``local'' energy updates. This is because for spin flips at some locations in the lattice one may need to recompute only some conditionals, but in general one has to recompute an extensive number of them 

Another limitation is that the RNN energy model is only able to make predictions for system sizes which are the same as those within the training set. Ideally, one would like to be able to observe an $L \times L$ system, learn something from it, and then make predictions for a larger $M \times M$ case.

\section{The EDNN Energy Model}

In previous work~\cite{ednn}, we showed that with the proper construction, neural networks have the ability to directly learn the locality length-scale, $l$, of operators such as the Hamiltonian. By locality length-scale, we mean the amount of information in the neighbourhood of a focus region, $f$, which is necessary to compute extensive properties. Magnetization, for example, is a fully local ($l=0$) operator. It is possible to divide the task of computing magnetism for every site in the system, record the value, and sum all at the end (since it is an extensive quantity). For a nearest neighbour spin model, additionally context, $c$, is needed in order to determine site energies, i.e. the values of the spins in the neighbourhood. Using an extensive deep neural network (EDNN), these locality scales can be learned directly from the data through hyperparameter optimization of $c$.

\begin{figure}[!h]
 \includegraphics[width=\columnwidth]{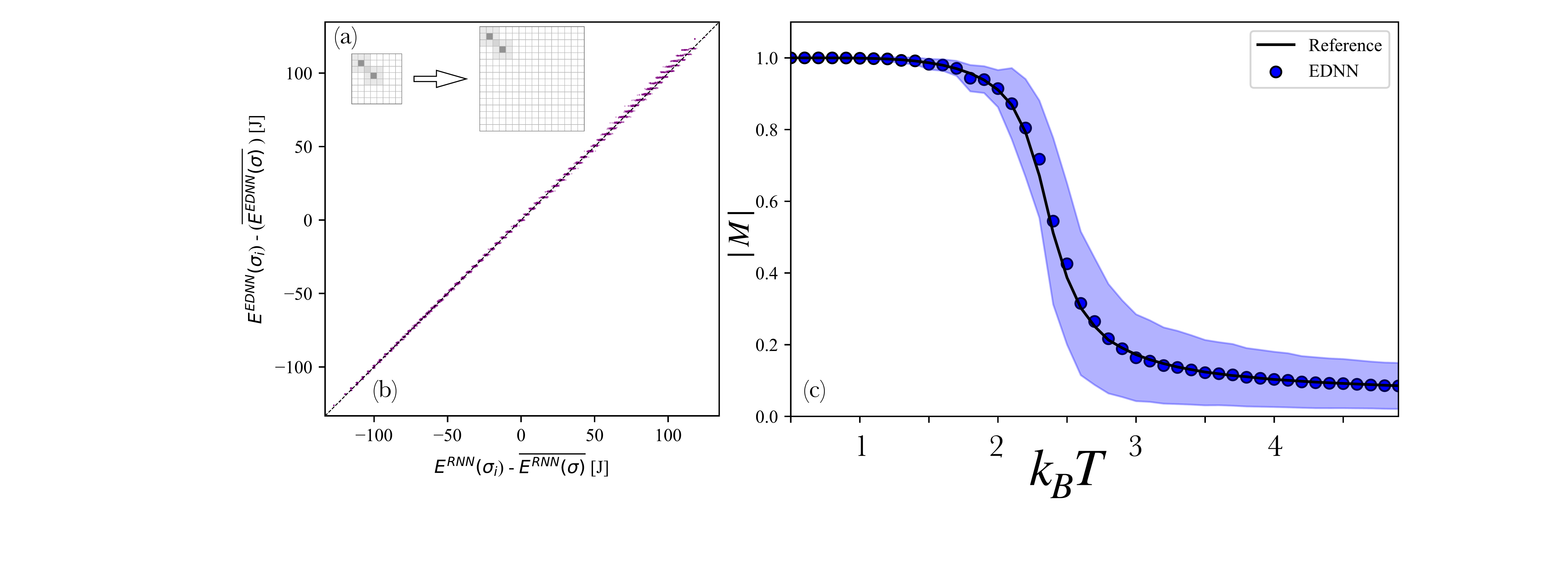}
 \caption{a) Schematic of an EDNN learning. By training on small scale examples of a problem ($8\times8$ Ising model), EDNN can learn the locality length-scale of an operator. b) Performance of an EDNN energy model trained on labels produced by the RNN energy model. b) Larger-scale ($16\times16$) simulations are now possible thanks to the interaction length-scale learning of the EDNN energy model. The model was trained on a data collected at a single temperature of a small lattice ($8\times8$) yet can be used to accurately predict the full phase behaviour of a larger lattice across all temperatures. This constitutes fully transferrable learning across conditions and length-scales. The blue shaded area indicates the variance observed in our numerical simulations.\label{fig:ednn_larger}}
\end{figure}

Initially, our motivation for using an EDNN in this investigation was to be able to learn from small scale systems (and $8\times8$ spin model in this case) and apply that learning to a larger system (e.g. $16\times16$). Training an EDNN with RNN energy model produced labels produces an EDNN energy model. As expected, the EDNN is able to take the small scale examples and transfer that learning to larger systems (Fig.~\ref{fig:ednn_larger} shows the performance of the EDNN energy model). Creating an EDNN energy model had a another unforeseen benefit, however.

By construction, EDNN topologies require that physical laws are the same everywhere. They are designed to learn a function which, when applied across a configuration, maps the sum of outputs to an extensive quantity such as the internal energy. Interestingly, in this case, we find that this physics-based network design requirement results in improved performance in predicting the underlying interactions even when the labels used in training have noise introduced by the imperfect RNN energy model.

\begin{figure}[!h]
 \includegraphics[width=\columnwidth]{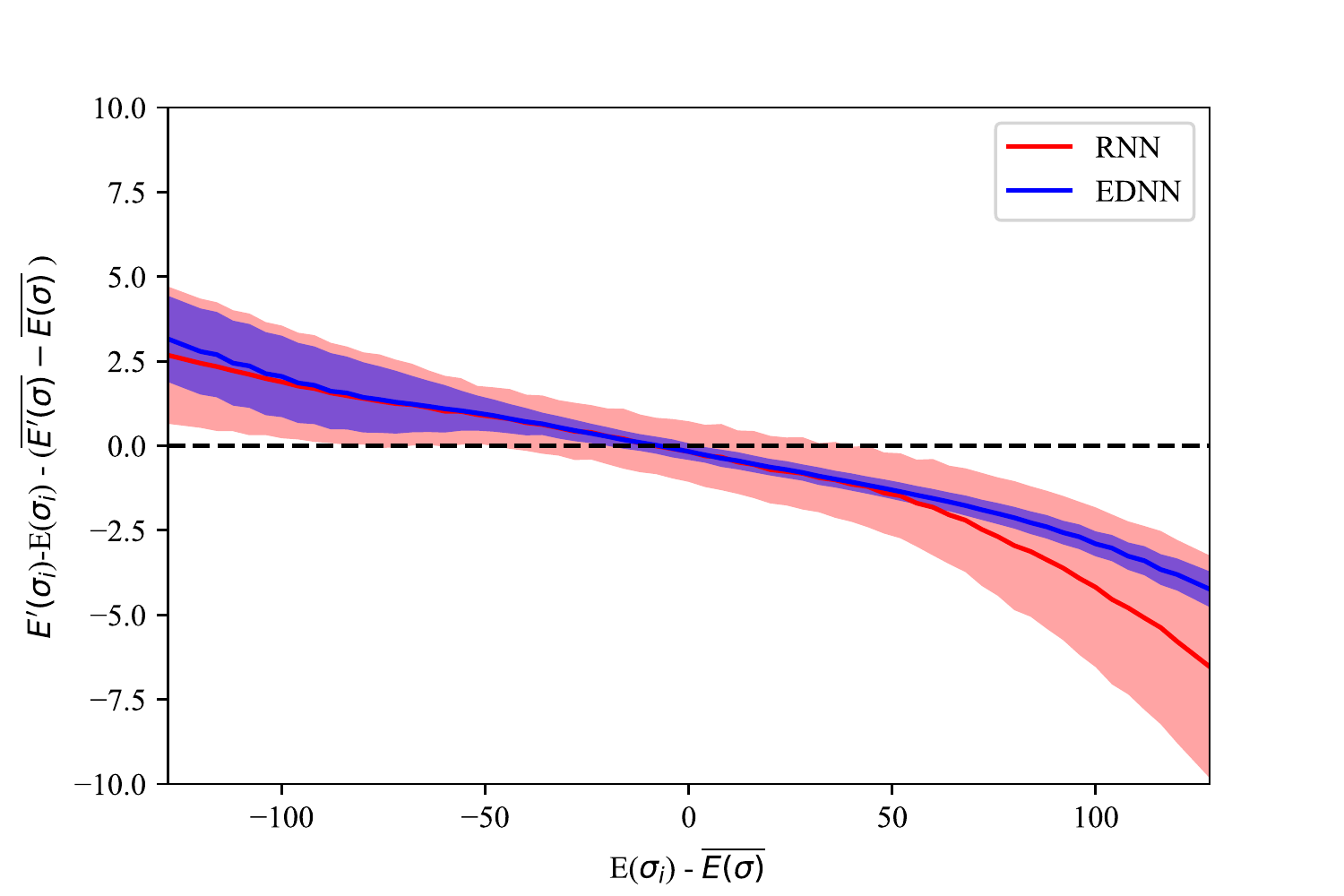}
 \caption{Comparison of RNN and EDNN energy models relative to true values of the energy over a wide range. The EDNN model has lower variance and lower RMSE relative to true data, even though it has never seen those labels. Results for the $8\times8$ Ising model are shown. Shared area indicates variance over a sample of 1000 configurations per energy window.\label{fig:ednn_vs_rnn}}
\end{figure}

As discussed above, we first train an RNN to predict energy differences $\Delta E$. In the second step, we train an EDNN to reproduce predictions from the RNN. The EDNN has never seen labels other than those estimated by the RNN. Despite this, when we compare EDNN predictions relative to the true values - it has better performance than the RNN itself. We suspect that the physics based construction of the EDNN enables it to see through noise introduced by the imperfect RNN and achieve a better estimate of the underlying operator (the root mean squared error, RMSE, of the RNN energy model for Ising is 2.06J compared with 1.47J for the EDNN, Fig.~\ref{fig:ednn_vs_rnn}). 
EDNN enforce the postulate of uniformity of physical law into our training procedure; our performance improves as a result.

\section{A general method}

\begin{figure}[t]
 \includegraphics[width=\columnwidth]{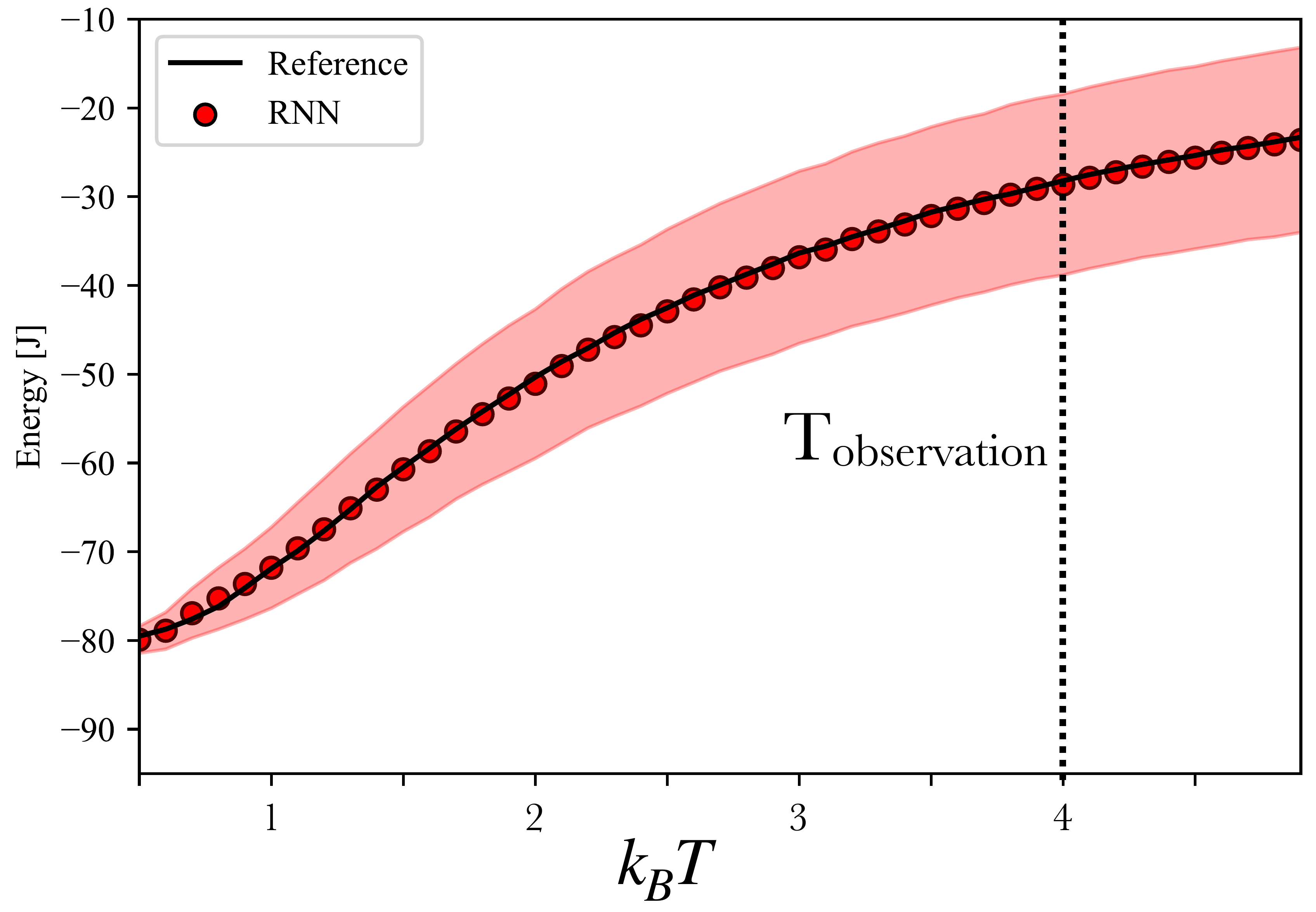}
 \caption{Predicting the finite temperature properties of a spin glass. Observations at of micro-states are made at a single fixed temperature, whereas predictions can be made across a wide range of conditions.\label{fig:spin_glass}}
\end{figure}

As a demonstration of the generality of DCL, we now consider the case of a spin-glass where interactions between neighbours are sampled from a random distribution (the Edwards-Anderson model). Even with a much more complex and rich Hamiltonian, the RNN is able to learn only from observations at a fixed temperature, yet can be used to make accurate predictions across a wide range of conditions. Again, only unlabelled observations at a single fixed temperature are required to determine the behaviour of the system under arbitrary conditions, Fig.~\ref{fig:spin_glass} (see S.I. for another example of random couplings).

By applying DCL to a spin-glass, we have demonstrated, for the first time, the ability to effectively learn the Hamiltonian operator directly without ever knowing its form, symmetries, or see direct labelled example. This is the first time such an inversion has been demonstrated for non-trivial systems.

\section{Conclusion}

With knowledge only of the experimental constraints (i.e. the statistical ensemble) and a statistically significant number of observations from only a \emph{single} set of thermodynamic conditions, distribution-consistent learning (DCL) is able to extract enough information about a physical system to fully predict its behaviour over a wide range of unseen conditions, across different length-scales, and through phase transitions.

DCL can predict the relative energy of micro-states at sufficient accuracy that it can be used to reproduce the energetic cost of excitations between states in a size consistent manner. The model consists of two deep neural network topologies which able to learn co-operatively. By using a combination of deep learning and physical constraints, we have shown that full transferability is possible. We expect that there will be many applications of this new method, including optical lattices, the growth of molecules on surfaces, among others.

\section{acknowledgments}
The authors acknowledge useful discussions with M. Fraser. IT and JC acknowledge support from NSERC. JC acknowledges support from, SHARCNET, Compute Canada, and the Canada CIFAR AI chair program. This research was supported in part by the National Science Foundation under Grant No. NSF PHY-1748958. Work carried out at NRC was done under the auspices of the AI4D Program. Work by SW was performed as part of a user project at the Molecular Foundry, Lawrence Berkeley National Laboratory, supported by the Office of Science, Office of Basic Energy Sciences, of the U.S. Department of Energy under Contract No. DE-AC02–05CH11231.

\section{Supplementary information}

Details of neural network topologies, training data, as well as additional tests and examples are outlined below.

\section{Neural network and training details}

We used 8$\times10^5$ configurations per spin for our training data (generated from two independent seeds). Given the high temperature of our observation, our simulations equilibrated rapidly. For test data, we sampled all possible energies (8000 samples from each). For the Ising model, this is $\approx 5\times10^5$. At the extremes, this is equivalent to reselecting certain configurations over and over again (consistent with what occurs physically). Test and train data were chosen such that they did not overlap. Our RNN are very simple, consisting of a single gated recurrent unit~\cite{gru} (we also tried up to 4 stacked units, which gave only a modest improvement in our results). The size of the hidden state was between 378-512 neurons. All RNN models converged quickly - results reported here are based on only 30 epoch (learning rates between $2.5-5\times10^{-4}$ and batch sizes between 4000-8000). For all models, we used dropout rate of 0.9. All of our networks are implemented in TensorFlow and are available online along with training data (http://clean.energyscience.ca/codes).

For the EDNN, we used $192\times10^{4}$ random spin configurations for training data  and achieved a converged result within only 60 epochs for all models. The EDNN was built using a previously reported architecture~\cite{ednn} ($f=2,c=1$ and 2 fully connected layers with 32 and 64 neurons with respectively). Throughout, we used rectified linear units (reLU) as activation functions. Our goal was to use a simple and consistent set of parameters and training; it is very likely that there exist better choices of hyper-parameters than those presented here.

We also note that the RNN we have used here are very simple in form. Recently, attention mechanisms\cite{bahdanau2014neural, kim2017structured} have been shown to improve the performance of sequence models significantly (e.g. reducing the number of needed samples need to achieve fixed fidelity). We expect that more advanced sequence models, including attention only ``Transformer''\cite{transformer} networks and related models could also be of benefit here, particularly with experimental data.

\begin{figure}[t]
 \includegraphics[width=\columnwidth]{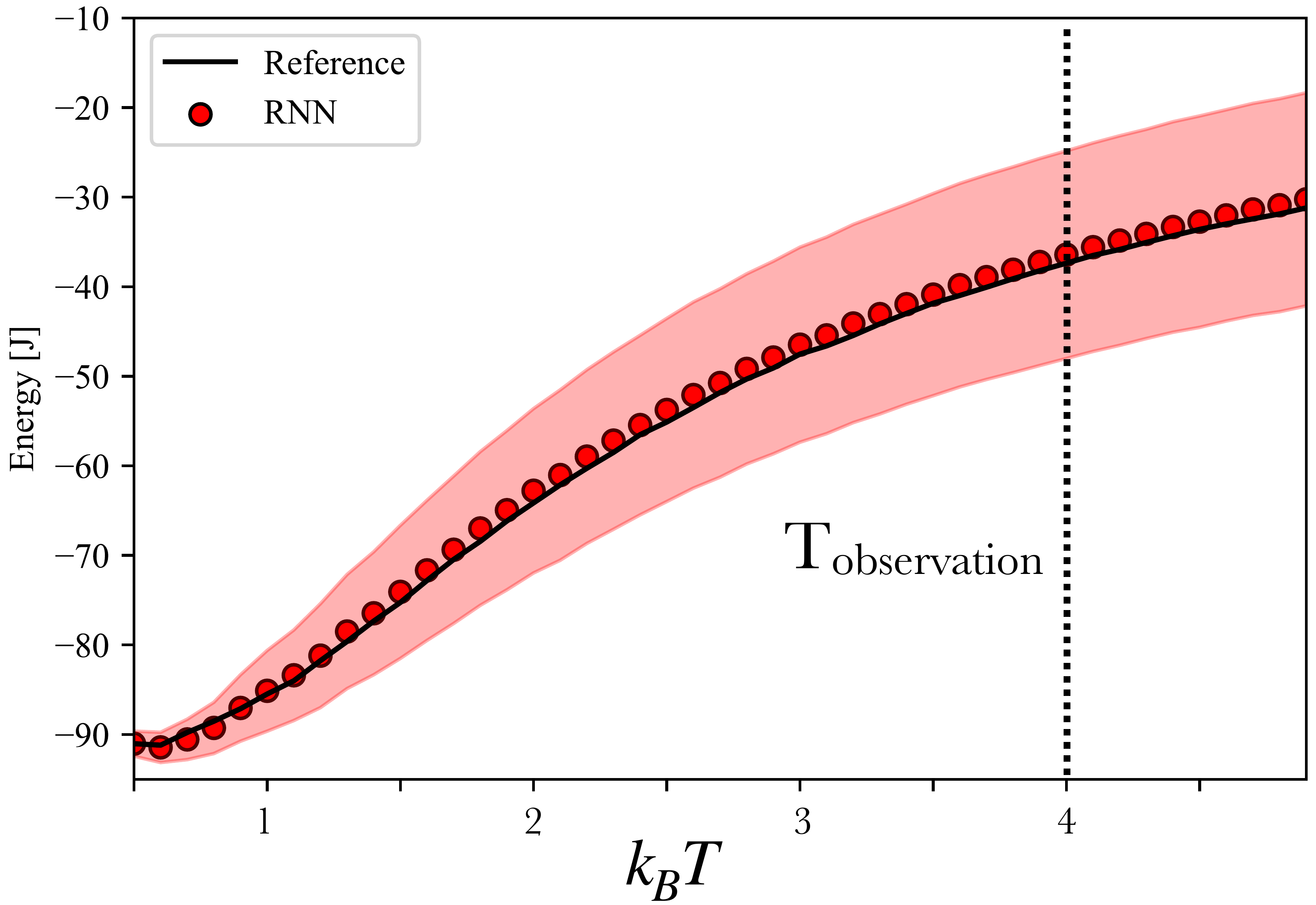}
 \caption{Supplementary Figure: Internal energy per spin as a function temperature for another spin-glass (i.e. a different set of random couplings).}
\end{figure}


\bibliography{wl_bib}

\end{document}